\documentclass[twocolumn,showpacs,amsmath,amssymb,prb]{revtex4}

\usepackage{amsmath}
\usepackage{amssymb}
\usepackage{graphicx}
\usepackage{dcolumn}
\usepackage{bm}

\def\be{\begin{eqnarray}}
\def\ee{\end{eqnarray}}
\def\ba{\begin{array}}
\def\ea{\end{array}}
\def\nn{\nonumber}

\begin{document}


\title{Realistic modelling of quantum point contacts subject to high magnetic fields and with current bias at out of linear response regime}

\author{S. Arslan$^1$} %
\author{E. Cicek$^2$}
\author{D. Eksi$^2$} %
\author{S. Aktas$^2$} %
\author{A. Weichselbaum$^1$} %
\author{A. Siddiki$^1$} %
\address{$^1$Physics Department, Arnold Sommerfeld Center for
Theoretical Physics, and
Center for NanoScience, \\
Ludwig-Maximilans-Universit\"at, Theresienstrasse 37, 80333
Munich, Germany}
\address{$^2$Trakya University, Faculty of Arts and Sciences, Department of Physics, 22030 Edirne, Turkey}


\begin{abstract}
The electron and current density distributions in the close
proximity of quantum point contacts (QPCs) are investigated. A
three dimensional Poisson equation is solved self-consistently to
obtain the electron density and potential profile in the absence
of an external magnetic field for gate and etching defined
devices. We observe the surface charges and their apparent effect
on the confinement potential, when considering the (deeply) etched
QPCs. In the presence of an external magnetic field, we
investigate the formation of the incompressible strips and their
influence on the current distribution both in the linear response
and out of linear response regime. A spatial asymmetry of the
current carrying incompressible strips, induced by the large
source drain voltages, is reported for such devices in the
non-linear regime.
\end{abstract}
\maketitle

\section{\label{sec:1} Introduction}
The new era of quantum information processing attracts an
increasing interest to investigate the intrinsic properties of
small-scale electronic devices. One of the most interesting of
such devices is the so called quantum point contacts (QPCs), where
a quantized current is transmitted through it under certain
conditions~\cite{Wees88:848,Whraham88:209}. They are constructed
on two dimensional electron systems (2DES) either by inducing
electrostatic potential on the plane of 2DES and/or by chemically
etching the structure. The essential physics is that, the small
size of the constraint creates quantized energy levels in one
dimension (perpendicular to the current direction) therefore,
transport takes place depending on whether the energy of the
electron coincides with this quantized energy or not. In the ideal
case at low bias voltages, if the energy of the electron is
smaller then the lowest eigen-energy of the constraint, no current
can pass through the QPC. Otherwise, only a certain integer number
of levels (channels) are involved, therefore conductance is
quantized~\cite{Davies}. Beyond being a useful play ground for the
basic quantum mechanical applications many other interesting
features are reported in the literature such as the 0.7
conductance
anomaly~\cite{Thomas96:135,Kristensen98:180,Kristensen00:10950},
which became a paradigm since then. Another adjustable parameter
which induces quantization on the 2DES is the magnetic field $B$
applied perpendicularly to the system. Such an external field
changes not only the density of states (DOS) profile of the 2DES
but also the screening properties of the system drastically. The
interesting physics dictated by this quantization is observed as
the quantum Hall effects~\cite{vKlitzing80:494,FQHE}. Recent
theoretical investigations point out the importance of the
electron-electron interactions in explaining the integer quantum
Hall effect~\cite{siddiki2004,afifPHYSEspin}, believed to be
irrelevant in the early days of this field~\cite{Kramer03:172},
which we discuss briefly in this work. The basic idea behind the
inclusion of the interaction is as follows: due to the
perpendicular magnetic field the energy spectrum is discrete,
known as the Landau levels (LLs), and is given by
$E_n=\hbar\omega_c(n+1/2)$ where $n$ is a positive integer and
cyclotron energy is defined as $\hbar\omega_c=\hbar eB/m^*$, with
effective electron mass $m^*=0.067*m_e$. Taking into account the
finite size of the sample, i.e. the confinement potential, and the
mutual interaction (Hartree) potential, one obtains the total
potential. In the next step for a fixed average electron density
one calculates the resulting electron density distribution and
from this distribution re-calculates the potential distribution
iteratively. This self-consistent calculation ends in formation of
the \emph{compressible} and \emph{incompressible} regions. In a
situation where Fermi energy $E_F$ is pinned to one of the LLs,
then the system is compressible. Otherwise, if $E_F$ falls in
between two consecutive LLs, the system is known to be
incompressible and, since there are no available states at the
$E_F$ for electrons to be redistributed, screening is poor.
However, within these incompressible regions the resistivity
vanishes, hence all the applied current is confined to these
regions. We will be discussing the details of this model in
Sec.\ref{sec:4}.

Most recently, the experiments performed at the 2DES including the
QPCs, in the presence of an external magnetic field, manifested
peculiar
results~\cite{Heiblum03:415,Tobias07:464,Roddaro05:156804}. In the
first set of experiments electron interference, such as
Mach-Zehnder
(MZ)~\cite{Heiblum03:415,Neder06:016804,Litvin07:033315,Roche07:161309,Litvin:2008arXiv,Roche07:QHP}
and Aharonov-Bohm (AB)~\cite{Goldman05:155313}, was investigated.
The MZ interference experiments exhibit a novel and yet
unexplained behavior, regarding the interference contrast
(visibility) at the interferometer in the nonlinear transport
regime (finite transport voltage). As a function of voltage, the
visibility displayed oscillations whose period was found to be
independent of the path lengths of the interferometer, in striking
contrast to any straightforward theoretical model (e.g. using
Landauer-B\"uttiker edge states (LBES)~\cite{Buettiker86:1761}).
In particular, a new energy scale, of order of $\mu$eV, emerged,
determining the periodicity of the pattern. This unexpected
behavior of interfering electrons is believed to be related to
electron-electron interactions~\cite{Heiblum03:415}. The most
satisfying model existing in the literature is proposed by I. P.
Levkivskyi and E. V. Sukhorukov~\cite{EugenMZI:08}. However, other
schemes also including interactions are
present~\cite{Chalker07:MZI} and models, which consider a
non-Gaussian noise as a source of the visibility
oscillations~\cite{Neder07:112}, without interaction. The novel
magnetic focusing experiments concerning QPCs has revealed the
scattering processes taking place near these
devices\cite{Tobias07:464}. It was explicitly shown that, the
experimental realization of the sample and the device itself
strongly effects the transport properties. It is reported that,
the potential profile generated by the donors (impurities) and the
gates deviates strongly from the ideal "point" contact. Even in
each cool down process, since the impurity distribution changes,
the quantum interference fringes differ considerably. Hence a
realistic modelling of a QPC is desirable, which we partially
attack in this work. Another interesting set of experiments within
the integer quantum Hall regime is conducted by S. Roddaro
et.al\cite{Roddaro05:156804}, where the transmission is
investigated as a function of the gate bias. The findings show
that current transmission strongly deviates from the expected
chiral Luttinger liquid~\cite{Dassarma} behavior, since the
transmission is either enhanced or suppressed by changing the gate
bias. This effect was attributed to the particle hole symmetry of
the Luttinger liquid and is discussed in detail in
Ref.[\onlinecite{Lal07:condmat}]. However, the explicit treatment
of the QPC was left unresolved. Since the essential physics can be
still governed by considering a finite size QPC opening, therefore
assuming formation of a (integer) filling region sufficient.

The theoretical investigation of QPCs covers a wide variety of
approaches, which can be grouped into two: i) the models that
include electron-electron interactions and ii) the models that do
not. At the very simple model in describing the QPCs, one
considers a potential barrier perpendicular to the current
direction quantizing the energy levels. Therefore the electrons
are considered to be plane waves before they reach to the QPC and
transmission and reflection coefficients are calculated from this
potential profile. A better (2D) approximation is to model the
QPCs with well defined
functions\cite{Chklovskii93:12605,Meir02:196802}, such as ellipses
which lead to analytic solutions for the energy eigenfunctions and
energies. About a decade ago J. Davies and his co-workers
developed the "frozen charge" model~\cite{Davies94:4800} to
calculate the potential profile induced by the gates defining the
QPC. This approach is still one of the most used technique to
obtain the potential profile, however, it is not self-consistent
and completely ignores the donors and surface charges. There
exists many theories which takes the potential profile from the
frozen charge model as an initial condition, and provides explicit
calculation schemes to obtain charge, current and potential
distributions~\cite{Igor06:075320,Rejec06:900,SiddikiMarquardt,enginPHYSEqpc,afifblau}.
One of the most complete scheme, even in the presence of an
external $B$ field, is the local spin density approximation (LSDA)
within the density functional theory (DFT)~\cite{kohnsham}. The
LSDA+DFT~\cite{Igor07:qpc1,Igor07:qpc2,Rejec06:900} approaches are
powerful to describe the essential physics of density distribution
and even 0.7 anomaly phenomenologically, however, the description
of the current distribution is still under debate. The scattering
problem through the QPCs is usually handled by the "wave packet"
formalism and is very successful in explaining the magnetic
focusing experiments. However, the potential profile is not
calculated self-consistently and therefore, the effect of the
incompressible strips resulting from electron-electron interaction
is not taken in to account.

Back to early days of the theories that account for electron
interactions, i.e. Chklovskii, Shklovskii and
Glazman~\cite{Chklovskii92:4026} (CSG) and Chklovskii, Matveev and
Shklovskii~\cite{Chklovskii93:12605} (CMS) models, the influence
of the formation of the incompressible strips has been
highlighted. In the CMS paper, it was even conjectured that,
\emph{'the ballistic conductance of the QPC in strong magnetic
field is given by the filling factor at the saddle point of the
electron density distribution multiplied by} $e^2/2\pi\hbar$',
which is quantized only if an incompressible strip (region)
resides at the saddle point. In one of the recent approaches, in
the presence of a strong $B$ field, the electron-electron
interaction is treated explicitly within the Thomas-Fermi
approximation (TFA) self-consistently, meanwhile the current
distribution is left unresolved~\cite{SiddikiMarquardt}. In this
model, similar to other approaches, the bare confinement potential
is obtained from the "frozen charge" approximation, which in turn
lead to discrepancies due to its non-self-consistent approach.
Here, we improve on this previous work in two main aspects: i) the
electrostatic potential is obtained self-consistently in 3D, which
allows us to treat also the etched structures ii) the current
distribution is calculated explicitly using the local version of
the Ohm's law, also in the out of linear response regime. We
organize our work as follows: In Sec.\ref{sec:2}, we briefly
describe the numerical scheme to calculate the potential profile
at $B=0$ following Ref.[\onlinecite{Weichselbaum03:056707}], which
is based on iterative solution of the Poisson equation in 3D. In
particular, we study the effect of different gate geometries and
focus on the comparison of the potential profiles of gate and etch
defined QPCs. The numerical scheme to calculate potential and
density profiles at finite temperature and magnetic field is
introduced in Sec.\ref{sec:3}. Here, we review the essential
ingredients of the TFA and discuss the limitations of our
approach. Sec.\ref{sec:4} is dedicated in investigating the
current distribution within the local Ohm's
law~\cite{Guven03:115327,siddiki2004,Bilayersiddiki06:}, where we
consider both the linear response (LR) and out of linear response
(OLR) regimes. In the OLR, we show that the large current bias
induces an asymmetric distribution of the incompressible strips,
due to the tilting of the Landau levels resulting from the
position dependent chemical potential. We conclude our work by a
summary.

\section{\label{sec:2} Electrostatics in 3D}
\begin{figure}
{\centering
\includegraphics[width=1.\linewidth]{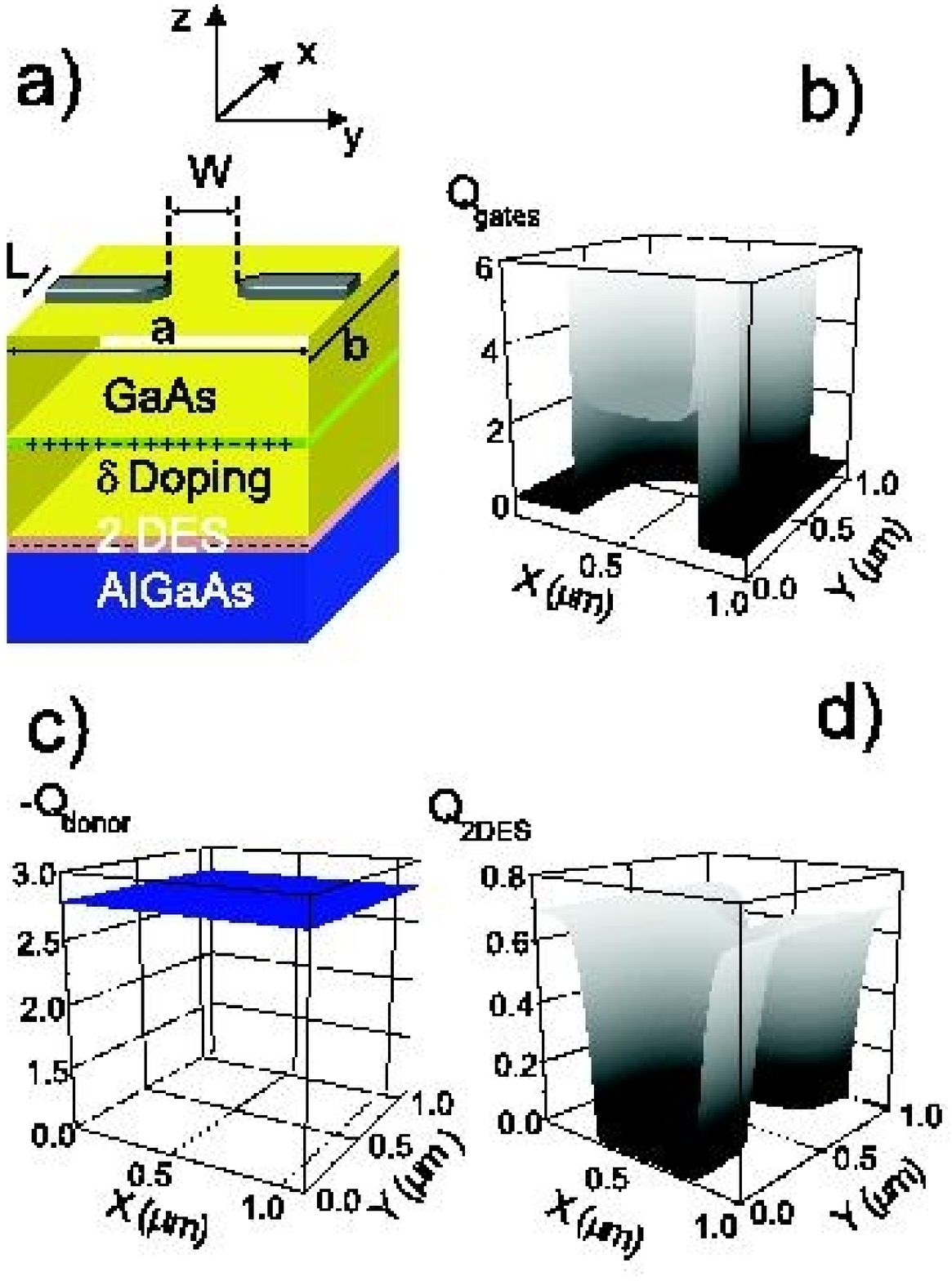}
\caption{ \label{fig:1} (Color online) (a) The Silicon doped
hetero-structure, the 2DES is formed at the interface of the
GaAs/AlGaAs (denoted by minus signs) and the metallic gates are
deposited on the surface. At zero gate bias, the electron density
is determined by the number of donors, which we chose to be
$4\times10^{16}$ cm$^{-2}$. Charge distribution at different
layers, at the gates (b), the dopant layer (c) and the 2DES (d).
It is clearly seen that not all the excess electrons are captured
by the 2DES, rather a significant amount is accumulated on the
surface. The electrostatic quantities are normalized with the
relevant scales, i.e. charge (density) is normalized with the
average electron density (e.g. ${\rm Q}_{\rm 2DES}(x,y)=n_{\rm
el}(x,y)/\bar{n}_{\rm el}$) and electrostatic potential (energy)
with the potential energy of a single electron.}}
\end{figure}

The realistic modelling of 2DES relies on solving the 3D Poisson
equation for given boundary conditions, set by the
hetero-structure (GaAs/AlGaAs in our calculations) and the gate
pattern, which describes the charge and potential distribution.
The hetero-structure, shown in Fig.\ref{fig:1}a, consists of
(metallic) surface gates (dark semi-elliptic regions on surface),
a thin donor layer (denoted by light thin layer and $\delta$
Silicon doping) which provides electrons to the 2DES and the 2DES
itself indicated by minus signs confined to a thin area. The 2DES
is formed at the interface of the hetero-structure. The average
electron density $\overline{n_{\rm el}}$ (and its spatial
distribution $n_{\rm el}(x,y)$) is dictated by the donor density
$\overline{n_{0}}$ and the metallic gates. Once the number of
donors and the gate voltage $V_G$ are fixed, the potential and
charge distribution of the system can be obtained by solving the
Poisson equation, self-consistently.

For typical nanoscale devices with many (or at least a few)
electrons in each of the electrostatically-defined regions, the
charge distribution and the major energy scales are described to a
good approximation by classical electrostatics. Due to the strong
electric fields generated by segregating charge in a 2DES, the
Coulomb energy is the dominant energy scale. In this sense, it is
desirable to have a self-consistent electrostatic description of
the system if one expects a good quantitative description thereof.

For solving the electrostatics of the system in three dimensions
we used a code developed and successfully applied in previous
studies~\cite{Andreas:06,Weichselbaum03:056707}. It is based on a
4$^{th}$ order algorithm operating on a square grid. The code
allows flexible implementation of many boundary conditions
relevant for nanoscale electrostatics: standard  boundaries such
as conducting regions at constant voltage (potential gates), of
constant charge (large quantum dots) or charge density (doping),
but also boundaries such as a depletable 2DEG, dielectric
boundaries and surfaces of semiconductors with the Fermi energy
pinned due to surface charges. Since the calculation is
constrained to a finite volume of space including the surface of
the sample, the outer boundary is considered open and is also
obtained self-consistently along with the rest of the calculation.

Overall the code provides a reliable description of the potential
landscape and thus the electric field as well as the charge
distribution for the sample under consideration.

As an illustrating example in Fig.\ref{fig:1} we show the
hetero-structure under investigation together with the charge
distributions at different layers. Area of the unit cell is
$1.5\times1.5$ $\mu$m$^2$, whereas the hight is chosen to be $156$
nm. The donor and the electron layer lies $43$ nm and $56$ nm
below the surface, respectively. The metallic gates are deposited
on the surface of the structure and are biased with $-1.7$ V and a
homogeneous donor distribution is assumed, Fig.~\ref{fig:1}c. The
induced charge distribution on the metallic gates exhibits
apparent inhomogeneities shown in Fig.\ref{fig:1}b. The electrons
are accumulated near the boundaries of the gates, a well known
behavior for metallic boundary conditions. However, here we
provide the explicit distribution which strongly differs from the
previously used "frozen charge" model where only a constant
potential profile is assumed. The influence of these induced
charges become more important when considering an external $B$
field. Since, the steepness of the external potential profile
determines the effective widths of the current carrying
incompressible strips. We should note that, our self-consistent
model enables us also to handle the (side) surface charges which
becomes important when considering chemical etching. In the
following part we investigate systematically, the effects of the
gate voltage and the device geometry on the electrostatic
quantities.
\begin{figure}
{\centering
\includegraphics[width=1.\linewidth]{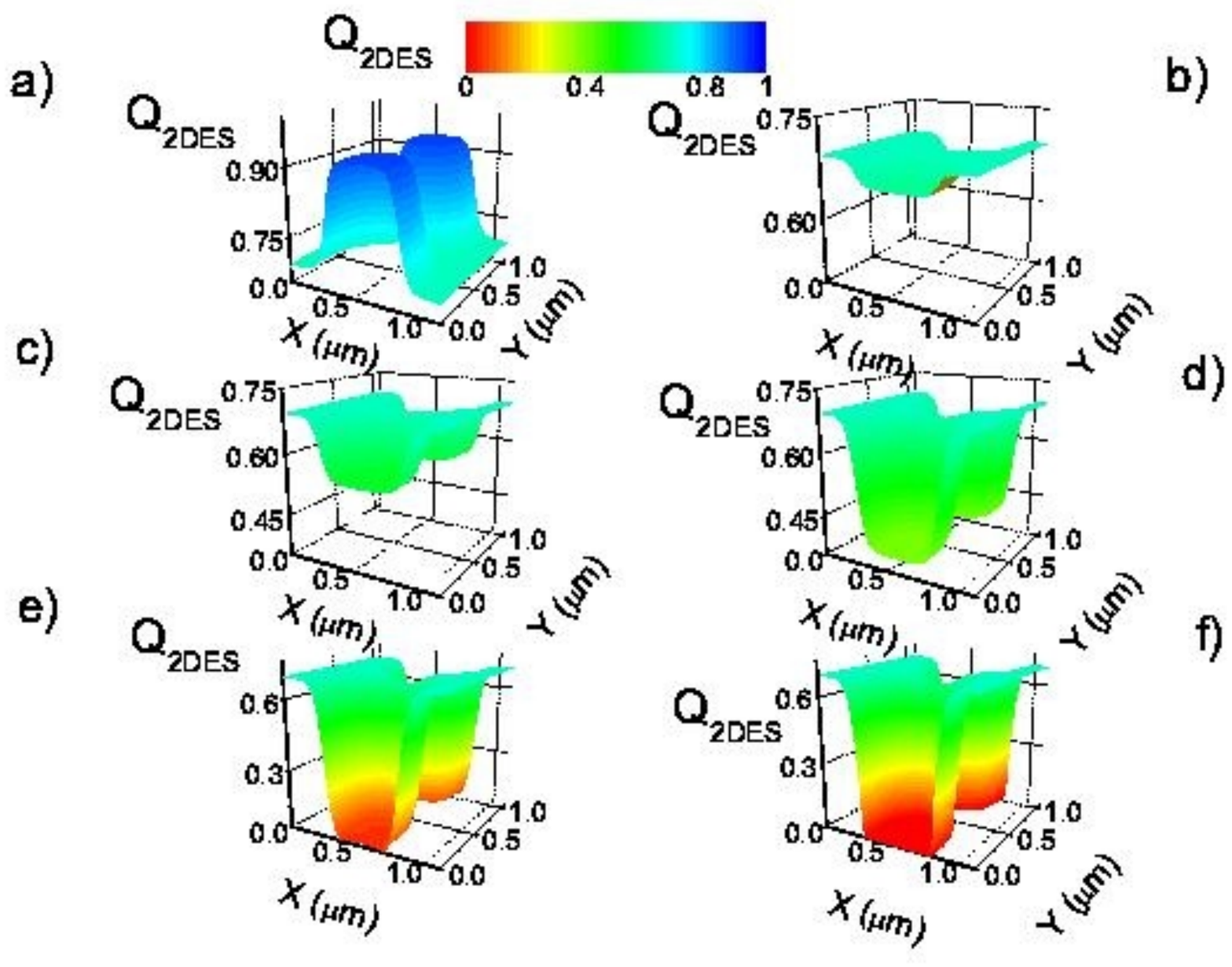}
\caption{ \label{fig:2} (Color online) Spatial distribution of the
electrons as a function of the gate voltage. At zero bias (a) more
electrons are populated under the gate which changes till
depletion starts (b-d), where the gate voltage is set to be (b)
-0.3V (c) -0.7V (d) -1.0 V. Almost no electrons are left beyond
-1.5V applied to the gates, (e) -1.5V (f) -2.0 V.}}
\end{figure}
\subsection{Gate defined QPCs}
In this subsection we compare the electron density profiles
calculated for different QPC geometries applying various bias
voltages, which exhibits strong non-linear behavior in contrast to
many models used in the literature. We start our discussion with a
rather smooth configuration, where the distance between the gates
($W$) is chosen to be $200$ nm (see Fig.\ref{fig:1}a). In Fig.
\ref{fig:2} we show the cross section of the electron density
profile for different gate voltages. Interestingly, at $V_G=0$ we
see that more electrons are residing beneath the gates. This
effect is due to inhomogeneous (induced) charge distribution at
the metallic gates similar to the distribution shown in
Fig.\ref{fig:1}d. The induced charges are mostly accumulated near
the gate boundaries, whereas the center of the gates has almost a
constant charge profile. Increasing $V_G$ to -0.3 Volts, already
starts to depopulate electrons under the gates and the
depopulation rate remains linear to the applied gate potential
until the 2DES becomes depleted. In the $[-0.5,-1.2]$ Volt
interval, the density distribution changes relatively smooth,
since the electrons can still screen the external potential quite
well. It is important to recall that, in the absence of an
external $B$ field, the DOS of a 2DES is just a constant $D_0$ ($=
m^*/\pi \hbar^2=2.83\times 10^{10}$ meV$^{-1}$ cm$^{-2}$ for
GaAs), which is set by the sample properties, therefore screening
is nearly perfect. Whereas this changes considerably when the
electrons are depleted under the gate. This is observed by the
strong drop of the potential when the depletion bias is reached
around $V_G=-1.3$ V. A sudden variation appears at the potential
profile when the electrons are depleted beneath the gates ($y=350$
nm) where the $V_G$ becomes larger than -1.5 V. Therefore, the
simple picture describing the QPCs as a smooth function of the
applied gate voltage fails. In that picture it is assumed that the
Fermi energy of the system remains constant and the potential
profile, given by a well defined function, of the constriction is
simply shifted by the amount of potential applied to the gates.
Such a model is reasonable in the regime where the gate voltage is
small enough that no electrons are essentially depleted. However,
as mentioned above, when the barrier hight is larger than the
Fermi energy there exists no electrons to screen the external
potential and the potential distribution must be calculated
self-consistently.

\begin{figure}
{\centering
\includegraphics[width=1.0\linewidth]{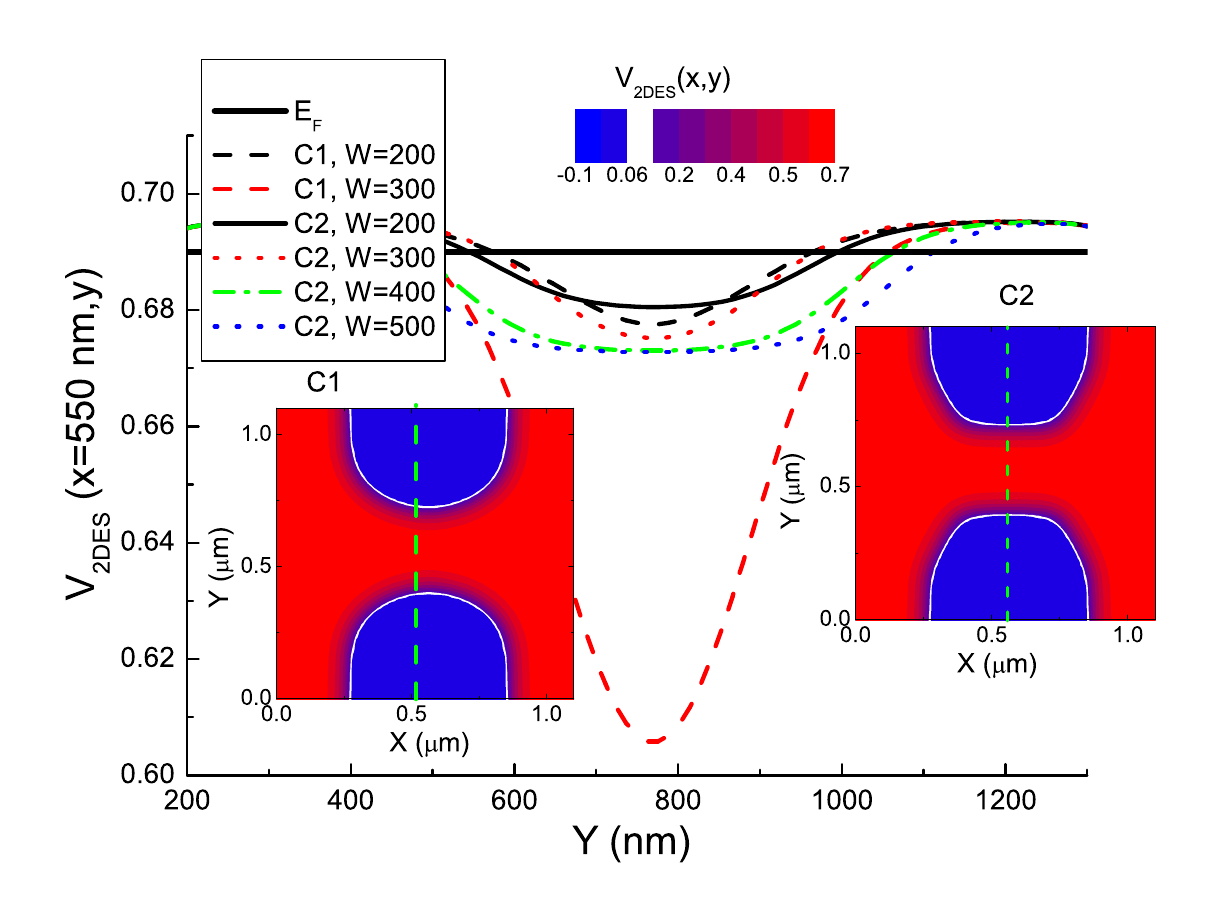}
\caption{ \label{fig:4} (Color online) Potential profile across
the QPCs for two different geometrical patterns. Insets depict the
smooth (C1) and the sharp edged (C2) patterns. Dark (blue) regions
are electron depleted, i.e. the local potential is larger than the
$E_F$, denoted by the solid thick horizontal line. White contours
denote the depletion boundary, where the width of the QPC is set
to be 200 nm and the curvature is changed. The dashed lines stand
for the first configuration and two different $W$ values ($=200$
nm (black) and $300$ nm (red)), whereas for the sharper
configuration four $W$ values are selected.}}
\end{figure}

Another adjustable parameter which can be accessed experimentally
is the geometry of the structure. Of course, in the simplistic
models describing QPCs this does not play an important role, since
the constriction is assumed to be isotropic in the current
direction, in contrast to the experimental findings. It is well
known that, the shape of the QPCs, as well as the cooling and
biassing procedure, is important when measuring interference or
magnetic focusing. In Fig.\ref{fig:4}, we compare two different
gate patterns considering typical gate separations $W$ for a fixed
gate voltage, $V_{G}=-2.0$ V. The smooth configuration (C1),
exhibits a stronger non-linear behavior when $W$ is changed from
$200$ nm (black dashed line) to $300$ nm (red dashed line). This
relies on the fact that, at the first order approximation, the
screening is better when more electrons are accumulated at the
opening of the QPC. However, since the screened potential $V_{{\rm
scr}}(x,y)$ can be obtained from the external potential $V_{\rm
ext}(x,y)$ via the Thomas-Fermi screening, \be V_{ {\rm
scr}}^{q}=V_{{\rm ext}}^{q}/\epsilon(q), \label{screen1} \ee where
$\epsilon(q)=1+2/(a^*_B|q|)$, is the Thomas-Fermi dielectric
function with $q$ being the momentum and $a^*_B(=9.8$ nm for GaAs)
the effective Bohr radius, the long range fluctuations (large
$q$), compared to $a^*_B$, are less screened whereas the short
range fluctuations are predominantly screened. Considering the
sharp configuration (C2) this observation becomes more evident,
since the potential profile across the QPC varies smoother than of
the configuration (C1), when varying $W$. From an experimentalists
point of view, therefore, drawing shaper QPC structures by
electron beam lithography may lead to a better (linear) control of
the potential profile which is closer to the ideal potential
profile. This is certainly in contrast to what we would expect
from an non-interacting model, however, it is known by the
experimentalists that defining the QPCs with sharper edges
increase the quality of the visibility signal~\cite{Neder06:priv}.
For the C2 configuration we also observe that, the potential
profile becomes almost insensitive to the width $W$ above 300 nm,
which coincides with our previous finding of better screening of
the long range fluctuations. It is worth to note that, our
calculation scheme is beyond the simple Thomas-Fermi screening
scheme in obtaining the bare confinement
potential~\cite{SiddikiMarquardt}. It fully takes into account the
interaction effects, however, does not include any quantum
mechanical effects. In a better approximation, of course, one
should also solve the Schr\"odinger equation self-consistently in
3D. This procedure is known to be costly in terms of computational
cost even only if the 2DES is treated quantum mechanically. Since
we are interested in either zero or very strong magnetic fields,
representing electron as a point charge is still a reasonable and
valid approximation. We will discuss the justification of this
assumption in the presence of an external magnetic field in more
detail in Sec.~\ref{sec:3}, where we also discuss the limitations
of our model.

The different configurations of the gate patterns are also
important when investigating the scattering processes by magnetic
focusing experiments~\cite{Tobias07:464}. It is apparent that the
scattering patterns of the electron waves will not only depend on
the impurity distribution but also on the structure of the QPC's.
We expect that for the sharper defined QPC (C2) the scattering
should depend weakly on the QPC opening, since the width does not
change along the constraint. Whereas, for C1 configuration small
changes at $W$ should affect the scattered waves drastically.
Another comment on the experimental setups is to the
Roddaro~\cite{Roddaro05:156804} experiments, since the formation
of (stable) integer filling region is important to
explain~\cite{Lal07:condmat} the findings, we believe that C2 type
configurations would be leading to a better resolution of the
transmission amplitudes. The choice of the structure and the width
apparently depends on the experimental interest, which is believed
to be irrelevant when modelling QPCs as ideal point contacts, and
the QPC's are not only defined by gates but alternatively also by
chemical etching. The gated structures are of course more
controllable, however, at high gate voltages required for
depletion, electrical sparks can occur, therefore the structure
can even be destroyed. In such situations etching defined QPCs are
preferred, although without further gates one loses the full
control of the potential profile. In the next section, we will
compare the potential profiles of etched and gate defined QPC's,
to show that in some cases etch defined QPC's may be more useful
to obtain steeper potential profiles.
\subsection{Etching versus gating}
\begin{figure}
{\centering
\includegraphics[width=1.\linewidth]{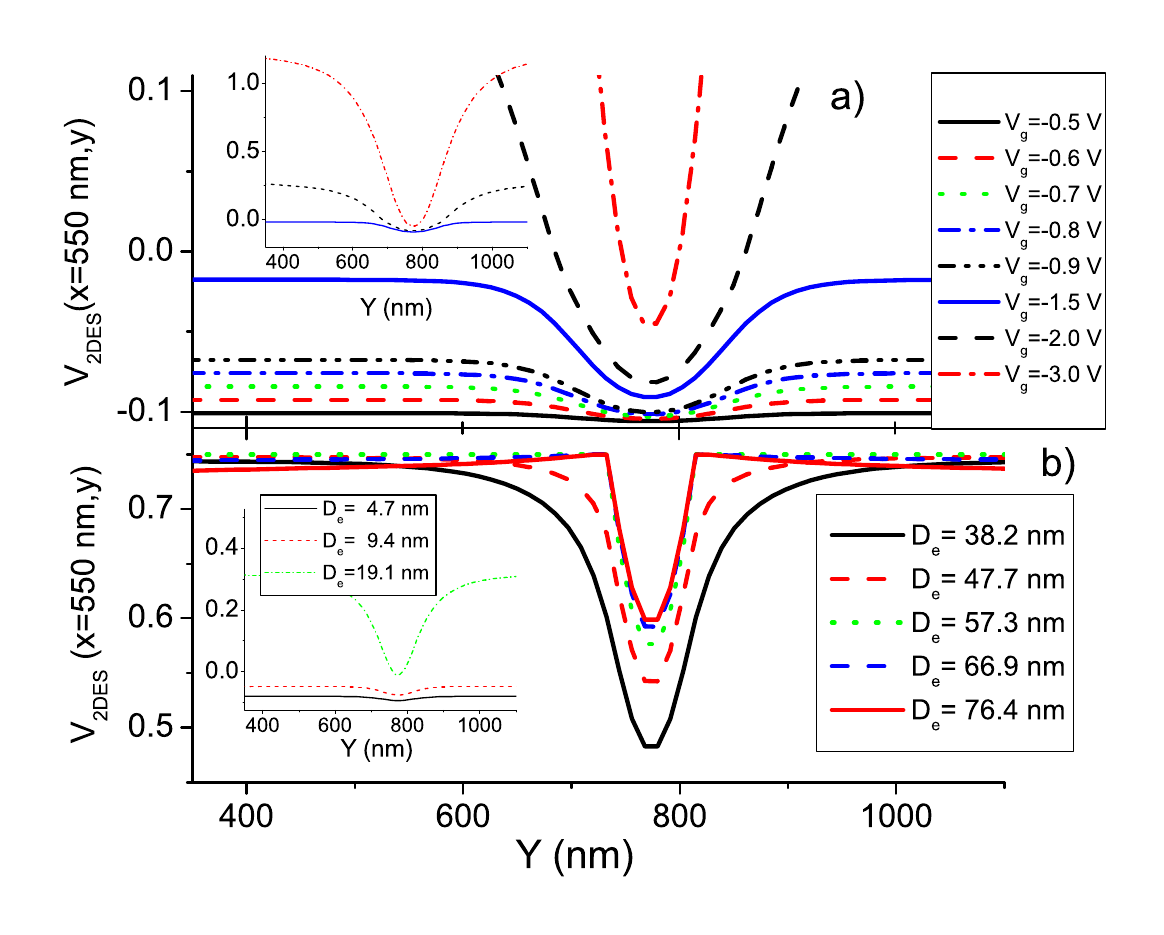}
\caption{ \label{fig:5} Potential cross-section of (a) gated and
respectively (b) etched QPC's at selected gate biases and etching
depths for C1 at a fixed $W=150$ nm. Insets focus on the high bias
or shallow etching profiles}}
\end{figure}
For simple calculation purposes, QPCs are modelled either as a
finite potential well or with a parabolic confinement potential,
perpendicular to the current direction. Starting from the early
experiments~\cite{Wees88:848}, usually the conductance is measured
as a function of the applied gate voltage which presents clear
quantized values. This quantization can be well explained by the
Landauer formula~\cite{Landauer81:91} \be G=\frac{e^2}{\pi
\hbar}\sum_{n,m=1}^{N_c}|t_{n,m}|^2, \label{eq:landauer}\ee where
ballistic transport is assumed to take place, i.e. the
transmission is given by $|t_{n,m}|^2=\delta_{n,m}$, and no
channel mixing is allowed. The (integer) number of channels $N_c$
is defined by the Fermi energy and the width of the constriction,
in general. The gate defined QPCs, at a first order approximation,
can be represented by parabolic or finite well potential profiles.
However, it is known that the chemical etching process creates
(side) surface charges, which in turn generates a steeper
potential profile at the edges of the sample. In this situation it
is apparent that, the confinement potential can not be assumed to
be parabolic, rather a steeper potential should be considered. In
this section we compare these two different constriction profiles,
namely the gated and the etched ones.

The self-consistent potential across the QPC at the center is
plotted versus the lateral coordinate in Fig.~\ref{fig:5}. The
2DES under the gates is depleted at the gate voltages larger than
$-1.5$ V, similarly when the depth of the etching is larger than
$47.7$ nm. In both cases, till the depletion is reached the
potential profile is varying rather smoothly and the depth of the
potential depends linearly on the applied gate bias or the depth.
This behavior is drastically changed when electrons are depleted,
the potential now strongly depends on the gate potential or depth.
Moreover, for the etched structure, potential becomes very steep
at the edge of the QPC when $D_{\rm e}>47.7$ nm, i.e. is deeper
than the donor layer. The transition from linear behavior to
non-linear behavior is simply due to the significant change of the
screening properties of the system. Once the electrons are
depleted, the external potential can no longer be screened,
therefore the amplitude of the self-consistent potential increases
by a large amount. Therefore, screening calculations based on the
formula given in Eqn.~(\ref{screen1}) can not account for such
situations, where the dielectric function is not aware of the
Fermi energy, i.e. the occupancy. A better approximation to this
approach is to consider a Linhard type dielectric function, which
also takes into account the Fermi momentum~\cite{Ando82:437}. It
is known that, such an improvement will also cover some of the
quantum mechanical aspects (such as the wave functions), which
brings extra oscillations to the potential
profile~\cite{Meir02:196802}. However, for our present interest we
neglect this correction knowing that the self-consistency of the
calculation scheme already takes into account the occupation and
the 1D electron density at the QPC satisfy the validity condition
$n_{\rm el}a_{B}^*>>1$ of the TFA~\cite{SiddikiMarquardt}.

\begin{figure}
{\centering
\includegraphics[width=1.\linewidth]{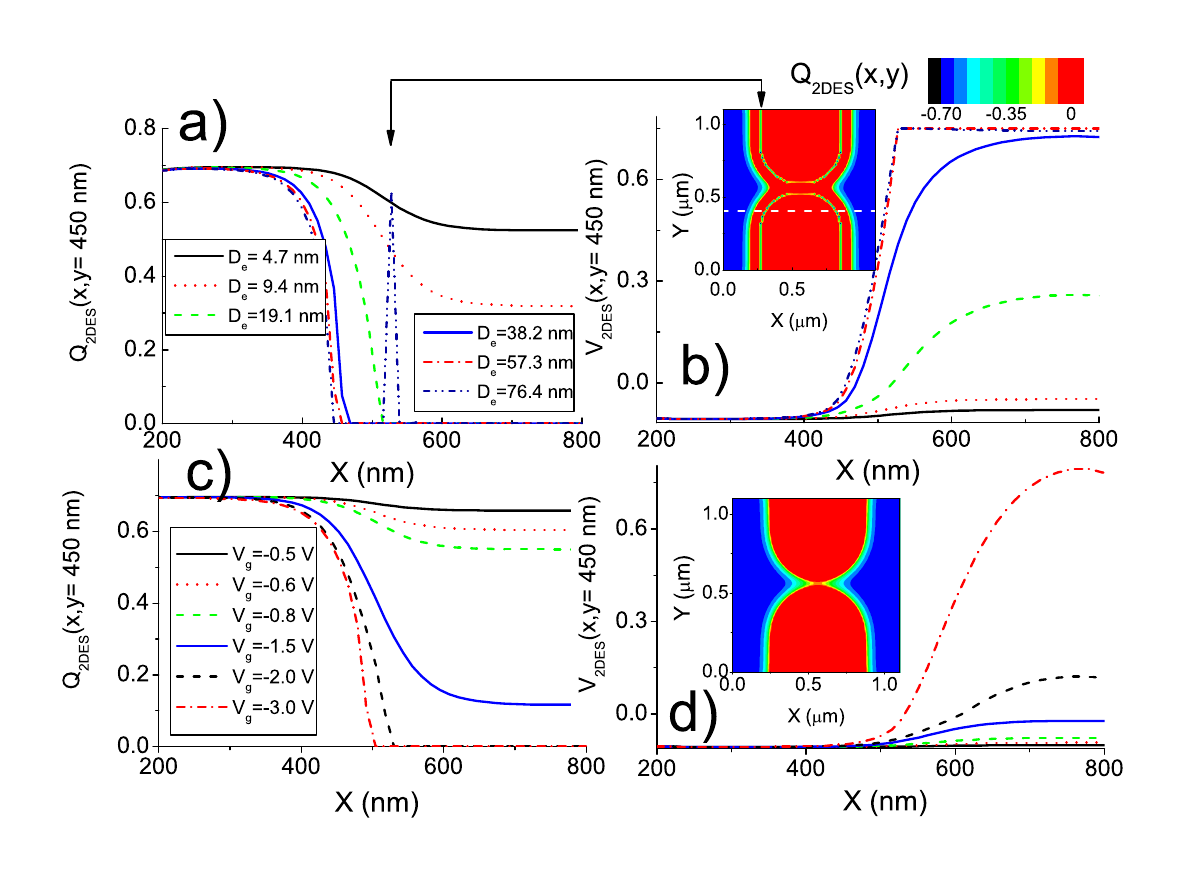}
\caption{ \label{fig:6} Spatial distribution of the electron
density (a,c) and screened potential (b,d) for etched (upper
panel) and gated samples (lower panel), $y=450$ nm.}}
\end{figure}

We summarize our findings in Fig.~\ref{fig:6}, where we show the
electron density (left) and potential profiles (right) for typical
gate biases $V_{\rm g}$ and etching depths $D_{\rm e}$ versus the
spatial coordinate. We choose a representative cross-section of
the obtained profiles along the current direction $x$, where the
$y$ coordinate is fixed at $450$ nm. Figure ~\ref{fig:6}a depicts
the density profile for selected depths of etching varying from
shallow ($D_{\rm e}=4.7-19.1$ nm) to deep ($D_{\rm e}=38.2-78.4$
nm). For the smallest two $D_{\rm e}$'s, the 2DES is not depleted
beneath the pattern and the density profile is rather smooth.
Depletion is observed when the depth is larger than $19.1$ nm,
however, until the etching depth reaches to the depth of 2DES
($\sim 60$ nm) we do not see the surface charges (the spike like
point, indicated by the arrow) at the level of the 2DES. The inset
of Fig.~\ref{fig:6}b shows the electron density distribution in a
color coded contour plot together with the corresponding potential
profile across the white (dashed) line. The thin (green) lines
contouring the depleted (red) region indicates the spatial
distribution of the surface charges. The potential is steeper
compared to that of the gated one (Fig.~\ref{fig:6}d) and the
profile does not show any considerable variation once the etching
depth reaches the plane of the 2DES. This behavior clearly
exhibits the uncontrollability of the etched samples, since the
corresponding potential profile obtained for the gated samples
vary slowly on the length scales of the Fermi wavelength, even if
the 2DES is completely depleted beneath the gates. Moreover, the
amplitude of the potential strongly depends on the applied gate
voltage. The slow variation of the potential is not the case for
the etched sample, for example consider the case when $D_{\rm e}$
is changed from 57.3 to 78.4 nm, and compare it with that of the
gated sample when voltage is changed from $-2.0$ V to $-3.0$ V.
For the gated sample the potential profile remains still smooth,
however, the spatial distribution of the electron density is
almost unchanged. Meanwhile, for the deeply etched sample both the
electron density and the potential profile are steep and the
steepness depends very weakly on the etching depth.

From the above discussion we conclude that, for the gated samples
the electron density distribution is weakly effected by the
applied gate potential when the depletion is once obtained,
meanwhile the potential is smooth and strongly depends on $V_{\rm
g}$. For the etched samples, potential profile becomes very steep
when the etching depth exceeds the depth of the 2DES, since the
(side) surface charges pin the Fermi level at the mesa surface to
the mid gap of GaAs forming a Schottky like
barrier~\cite{Siddiki03:125315}. We should also note that, at zero
bias, more electrons are populated under the gates, which is not
the case for the etched samples. As a rule of a thumb, when a
steeper potential profile is required one should consider chemical
etching where the etching depth is deeper than the electron layer
and one should keep in mind that biasing gates with high potential
does not necessarily imply that the electron density profile is
also changed considerably.

The outcome of the self-consistent solution of the 3D Poisson
equation considering QPCs at zero magnetic field is two fold: i)
the geometrical properties (i.e. considering C1 or C2 type
patterns) strongly change the potential landscape in the close
vicinity of the QPCs. We found that the smoother constrictions
with a larger width $W$ can be modelled better with the ideal
point contacts, i.e. parabolic confinement. The sharper
constrictions can be considered as finite well profiles, up to a
first order approximation and potential profile remains unchanged
when considering $W>300$ nm. ii) Due to surface pinning the etched
samples, generate steeper potential profiles and the density
profile remains unaffected once etching is deeper than the depth
of the 2DES. These numerical results, show a strong deviation from
the the widely used "ideal point contact" and "frozen charge"
models, proposing that depending on the experimental interest it
is important to reconsider the geometrical (C1, C2) and structural
(gated/etched) factors defining the QPC under investigation. As
final remark, the artifacts, such as local minima and maxima at
the potential and density profiles, resulting from the previous
non-self consistent Thomas-Fermi~\cite{SiddikiMarquardt} and
"frozen charge"
models~\cite{Davies94:4800,Igor07:qpc1,Rejec06:900} are resolved
by considering the 3D calculation scheme.
\section{Finite B Field\label{sec:3}}
The aim of this and the next section is to calculate, and compare,
the density and subsequently the current distributions, in the
close vicinity of the QPCs, within interacting and non-interacting
models in the presence of an external perpendicular magnetic
field. Here we take into account electron-electron interaction
within the Thomas-Fermi theory of screening also considering
strong magnetic fields using the potential profile calculated in
the previous section, as an initial configuration of the
landscape. We compare the spatial distribution of the
Landauer-B\"uttiker edge states (LB-ES) with the distribution of
the incompressible edge states, where the applied external current
is
confined~\cite{Chang90:871,Fogler94:1656,Guven03:115327,siddiki2004}.
Next we discuss the limitations of the TFA, and suggest
improvements on the calculation scheme based on i) quantum
mechanical considerations, such as the finite extension of the
wave functions, and ii) replacement of the global density of
states with the local one.

\subsection{Thomas-Fermi-Approximation (TFA)}
The enormous variety of the theories describing the density and
current distributions at the quantum Hall
systems~\cite{Laughlin81,Buettiker86:1761,Chang90:871,Chklovskii92:4026,Fogler94:1656,Guven03:115327,siddiki2004}
already show the challenge in giving a proper prescription to
these quantities. These theories can be grouped into two: the
current is carried either (i) by the compressible
regions~\cite{Chklovskii92:4026,Beenakker89:2020,Buettiker88:317}
or (ii) by the incompressible
regions~\cite{Chang90:871,Fogler94:1656,Komiyama96:558,Guven03:115327,siddiki2004}.
Moreover (and confusingly) these regions can reside at the
bulk~\cite{Laughlin81,Kramer03:172} or at the edge of the
sample~\cite{Buettiker88:317,Chang90:871,Chklovskii92:4026,Guven03:115327,siddiki2004,Akera06:},
depending on the model considered and the magnetic field
strength~\cite{siddiki2004}. For the sake of completeness, we
start with a generic Hamiltonian describing an electron subject to
high magnetic fields. \be H^{\sigma}=H_0+V^{\sigma}_{\rm
int}+V_{\rm ext}+V^{\sigma}_{\rm Z}, \label{eq:3}\ee where
$\sigma(=\pm 1/2)$ is the spin degree of freedom, $H_0$ the
kinetic part, $V_{\rm ext}$ and $V_{\rm int}$ the external and the
interaction potentials, respectively, and $V^{\sigma}_{\rm Z}$
Zeeman term~\cite{Igor06:075331}. Our first assumption is to
neglect the spin degeneracy, knowing that the effective band $g-$
factor for the GaAs/AlGaAs hetero-structures is a factor of four
less than the one of a free electron gas, and therefore Zeeman
splitting is much smaller than the Landau splitting ($\hbar
\omega_c$, with $\omega_c=eB/m^*$, i.e. $|g^*\mu_BB|/\hbar
\omega_c\approx 0.027$, where $\mu_B$ is the Bohr magneton.
However, Zeeman splitting can be as large as the Landau splitting
if exchange and correlation effects are taken into account at
significantly high magnetic fields, hence filling factor $\nu$
($=E_F/\hbar \omega_c $) one plateau can be observed
experimentally. On the other hand, for higher filling factors
$\nu>1$ exchange and correlation effects are assumed to be small
and Zeeman splitting is negligible. Thus one can consider
spin-less electrons in the magnetic field interval we are
interested in, which yields to the effective Hamiltonian, \be
H^{\rm eff}=H_0+V_{\rm ext}(x,y)+V_{\rm int}(x,y).
\label{eq:effecham}\ee The kinetic part, $H_0$, can be solved
analytically using the Landau gauge which yield plane wave
solutions in one direction ($y$) and Landau wavefunctions in the
other direction ($x$). Here, implicitly, a long Hall bar (ideally
infinite) is assumed, which is justified while the Fermi wave
length is ($\sim 30-40$ nm) much smaller than the sample length
under consideration ($L_y$ $\sim 1500$ nm). Then the
eigenfunctions of $H_0$ can be expressed as \be
\Phi_{n,k_y}(x,y)=\frac{1}{
\sqrt{2^n n! \sqrt{ \pi} l L_y}}\exp{(ik_y.y)}\times \nn \\
\exp{[-(\frac{x-X}{l})^2/2]}\times H_n(\frac{x-X}{l}), \ee where
$k_y$ is the quasi-continuous momentum in $y$ direction, $n$ the
Landau index, $X=-l^2k_y$ a center coordinate, and $H_n(\xi)$ the
$n$th order Hermite polynomial with the argument $\xi$, whereas
the eigen energies are \be E_{n}=\hbar \omega_c(n+1/2).\ee The
essence of the TFA relies on the fact that the potential profile
varies smoothly on the quantum mechanical length scales. Through
out this work we will only consider the  $6-8$ T interval and as a
rough estimation the extend of the wavefunctions in $x$ direction
(of the ground state, i.e. $\nu=2$), or the magnetic length
$l=\sqrt{\hbar/m \omega_c}$, will be always similar or less than
$10$ nm, therefore in almost all cases neglecting the finite
extend of the wavefunctions is still reasonable. However, we have
already seen that for the etched samples the potential is quite
steep and the results obtained from the TFA may be doubted, which
we will address in the next section.

At the moment let us consider a case that the condition of TFA
holds, i.e. the total potential $V_{\rm tot}(x,y)=V_{\rm
ext}(x,y)+V_{\rm int}(x,y)$ varies smoothly on the quantum
mechanical length scales and the sample is long enough (i.e.
$k_FL_y\gg1$). Then we can replace the wavefunctions in both
directions by wave packets centered at $X$, and at $y$ and the
center coordinate dependent eigenenergy $E_{n}(X)$ can be
approximated to $E_n+V_{\rm tot}(X,y)$. It follows that the
spatial distribution of the electron density within the TFA is
given by the expression
~\cite{Lier94:7757,SiddikiMarquardt,enginPHYSEqpc}, \be n_{\rm
el}(x,y)=\int D(E,(x,y))f(E+V_{\rm tot}(x,y)-\mu^*) dE
\label{eq:tfadensity}\ee with $D(E,(x,y))$ (local) density of
states, $f(E)=1/[\exp(E/k_bT)+1]$ the Fermi function, $\mu^*$ the
electrochemical potential, which is a constant in equilibrium
state, $k_B$ the Boltzmann constant, and $T$ the temperature. Once
the electron density is obtained, the interaction potential, i.e.
the Hartree potential, can be obtained from \be V_{\rm
int}(x,y)=\frac{2e^2}{\bar{\kappa}}\int_{A}K(x,y,x',y')n(x',y')dx'dy'
.\label{eq:tfapotential}\ee Here, $\bar{\kappa}$ is an average
dielectric constant ($=12.4$ for GaAs) and $K(x,y,x',y')$ is the
solution of the 2D Poisson equation satisfying the boundary
conditions dictated by the sample. The results reported in this
and the following section assume periodic boundary conditions,
where a closed form of the kernel $K(x,y,x',y')$ can be obtained
analytically~\cite{Morse-Feshbach53:1240}. Equations
(\ref{eq:tfadensity}) and (\ref{eq:tfapotential}) form a
self-consistent loop to obtain the potential and the density
profiles of a 2DES subject to high perpendicular magnetic field in
the absence of an external current at equilibrium, which has to be
solved iteratively using numerical methods. The computational
effort to calculate electron and potential profiles within the TFA
is much less than that of the full quantum mechanical calculation
procedures. The results of both agree quantitatively very well in
certain magnetic field intervals where the widths of the
incompressible strips $W_{\rm IS}$ (in which the potential changes
strongly) is larger than $l$. If $W_{\rm IS}\lesssim l$ condition
is reached, first of all the TFA becomes invalid and the
calculation of the electron density should include the finite
extend of the wavefunctions. The underestimation of the quantum
mechanical effects lead to existence of artificial incompressible
strips both in the non self-consistent electrostatic approximation
(NSCESA)~\cite{Chklovskii92:4026,Chklovskii93:12605} and
self-consistent TFA schemes~\cite{Oh97:13519,Siddiki03:125315}. In
fact, as early as the NSCESA, the self-consistent schemes which
also took into account finite extend of the wave functions already
pointed out the suppression of the incompressible strips in
certain magnetic field intervals~\cite{Suzuki93:2986} and also in
the recent reports~\cite{siddiki2004,Igor06:075320}. For a
systematic comparison of the calculated widths of the
incompressible strips within the TFA and the full Hartree
approximations, we suggest the reader to check Ref.
[\onlinecite{siddiki2004}], where a simpler quasi-Hartree scheme
is proposed to recover the artifacts arising from TFA.
\subsection{Corrections to the TFA}
Historically, the first implementation of the TFA, including
electron interactions, to quantum Hall systems goes back to the
seminal work by Chklovskii et.al~\cite{Chklovskii92:4026}. There
it was shown that within the electrostatic approximation the 2DES
is divided into two regions which have completely different
screening properties. In this model, a translation invariance is
assumed in the current ($y-$) direction. Due to finite widths of
the samples in the $x$ direction the electrostatic potential is
bent upwards at the edges of the sample, hence the Landau levels,
$E_{n}(X)=E_n+V(X)$. Inclusion of the Coulomb interaction and the
pinning of the Fermi energy to the Landau levels result in two
regions (strips): i) The Fermi energy is pinned to one of the
highly degenerate Landau levels, then the screening is perfect,
effective potential is completely flat (metal like) and electron
density varies spatially ii) The Fermi energy falls in between two
consecutive Landau levels, screening is poor, effective (screened)
potential varies (the amplitude of the variation is $\hbar
\omega_c$) and electron density is constant over this region. It
is apparent that, if the potential varies rapidly on the scale of
$l$, the TFA fails and the results become unreliable. This
condition is realized when considering narrow incompressible
strips having a width smaller than the magnetic length. Therefore,
one should include the effect of wave functions within these
narrow strips. One way is, of course, to do full Hartree
calculations. We already mentioned the challenges in the
computational effort. A simpler approach is to replace the delta
wave functions of the TFA with the unperturbed Landau wave
functions, i.e. quasi Hartree
approximation~\cite{siddiki2004}(QHA). The findings of the QHA is
shown to be more reasonable than of the TFA, which now also
includes the finite extend of the wave functions in the close
vicinity of the incompressible strips. Therefore, as an end
result, when the $W_{\rm IS}\lesssim l$ condition is reached the
incompressible strip disappear due to the overlap of the
neighboring wave functions. Based on this fact, in our
calculations in the following we will exclude the effects arising
from the artifacts of the TFA by considering a spatial averaging
of the electron density on the length scales smaller than $l$,
which is known to be relevant in simulating the quantum mechanical
effects~\cite{Siddiki:ijmp,siddikikraus:05,Bilayersiddiki06:}.

We should also make one more point clear that with the NSCESA
usually a gate defined quantum Hall bar is considered. It is more
common to define Hall bars by chemical etching and the edge
potential profile is much more steep compared to gated samples
which was shown in the previous section. Therefore, to fit the
predictions of this model concerning the widths and the positions
of the incompressible strips with the experimental data one has to
assume that~\cite{Ahlswede01:562} i) the 2DES and the gates are on
the same plane and ii) the gate voltage applied should be fixed to
the half of the mid gap of the GaAs, i.e. pinning of the Fermi
energy at the GaAs surface. In fact, after making these two
crucial assumptions the experimental findings of E.
Ahlswede~\cite{Ahlswede01:562} perfectly fits with the NSCESA.
However, the widths (and the existence) of the incompressible
strips strongly deviate from the predictions, since only the
innermost incompressible strip can be observed. We have argued
that, the widths of the incompressible strips strongly depend on
the slope of the potential, i.e. if the external potential is
steep the incompressible strips are narrow. Therefore one can
easily conclude that, since the widths of the outermost
incompressible strips become smaller than the magnetic length, the
outer incompressible strips, i.e. the ones close to the edge,
disappear and could not be observed. The overestimation of the
$W_{\rm IS}$ within the TFA becomes more severe when an external
current is imposed to the system, which we will discuss in the
next section. Before discussing the results of the relaxed TFA, we
want to touch another locally defined quantity, namely the DOS,
and comment on the implementation of the global DOS to our local
TFA.

In the absence of impurity scattering the DOS of an infinite
(spin-less) 2DES is given by the bare Landau DOS as \be D(E)=
\frac{1}{2 \pi l^2}\sum_n{\delta(E-E_{n})} \ee however, this DOS
is broadened by the scattering processes, which can be described
in self-consistent Born
approximation~\cite{Ando82:437,siddiki2004} accurately for short
range impurity potentials yielding a semi elliptic broadening. Of
course, other impurity models and scattering processes can also be
considered resulting in Gaussian or Lorentzian broadened Landau
DOS~\cite{Guven03:115327,deniz06}. In such descriptions of the DOS
broadening an infinite 2DES is assumed and the DOS is calculated
for impurity distributions averaged over all possible
configurations. Inserting this (global) DOS in
Eqn.~(\ref{eq:tfadensity}) can be justified again if the TFA
condition is satisfied. We have already shown that this condition
is violated when a narrow incompressible strip is formed where the
external potential is poorly screened. Therefore, the actual
distribution of the impurity potential becomes more effective at
these transparent regions. We should also note that the effect of
screening on the DOS of an infinite system has been investigated
in detail in Ref.~[\onlinecite{Cai86:3967}] and it has been shown
that, since the screening is poor within the incompressible
regions, the DOS becomes much broader than that of the
non-interacting case. Hence, the gap between two consecutive
Landau levels is narrower within the incompressible strips
compared to compressible strips. Moreover, recently it has been
shown that the (local) electric field within the sample also leads
to broadening of the (local) DOS~\cite{TobiasK06:h}. The idea is
basically that one calculates the Greens function for the given
potential profile, which is a function of the applied magnetic
field and external current, and obtains the local DOS from the
general expression \be
D(E,(x,y))=\sum_n{|\widetilde{\Phi}_{n,k_y}(x,y)|^2\delta(E-\widetilde{E}_{n,k_y})},\ee
where $\widetilde{\Phi}_{n,k_y}(x,y)$ is the $n$th eigenfunction
of the Hamiltonian given at Eqn.~(\ref{eq:effecham}) with the
eigenvalue $\widetilde{E}_{n,k_y}$. In our above discussion about
the formation of the compressible/incompressible strips we have
mentioned that the potential varies locally whenever an
incompressible strip is formed, where the variation is linear in
position up to a first order approximation. Now let us consider a
linear potential profile and re-obtain the local DOS (LDOS)
following~\cite{Kramer04:21} for the $k$th Landau level, \be
D_k(E)=\frac{1}{2^{k+1}k!\pi^{3/2}l^2\Gamma}e^{-E^2_k/\Gamma^2}[H_k(E_k/\Gamma)]^2
\label{eq:ldos}\ee with the level width parameter \be
\Gamma=\mathcal{E}_xl \ee where
$\mathcal{E}_x=\partial{V(x,y)}/\partial{x}$ is the electric field
in the $x$ direction and $E_k=E-\Gamma^2/(4\hbar
\omega_c)+(2k+1)\hbar \omega_c$. The immediate consequence of a
strong electric field in the $x$ direction is a broadening of the
LDOS, which happens at the incompressible strips. On the other
hand, since $\mathcal{E}_x$ vanishes at the compressible strips,
the bare Landau DOS is reconstructed from Eqn.~(\ref{eq:ldos}) in
the $\Gamma \rightarrow 0$ limit.

In summary, the TFA should be repaired when the widths of the
incompressible strips is comparable or less than the magnetic
field, since (i) the quantum mechanical wave functions have a
finite extend and do overlap with the neighboring ones (ii) the
LDOS are broadened where strong electric fields exist, (iii)
within the poor screening regions the (Landau) gap is smaller than
the ones of the nearly perfect screening regions. At the very
simple approximation this artefact of TFA is cured by a spatial
averaging over the Fermi wavelength resulting in nonexistence of
narrow incompressible strips. In the following an appropriate
averaging process will be applied.
\begin{figure}
{\centering
\includegraphics[width=1.\linewidth]{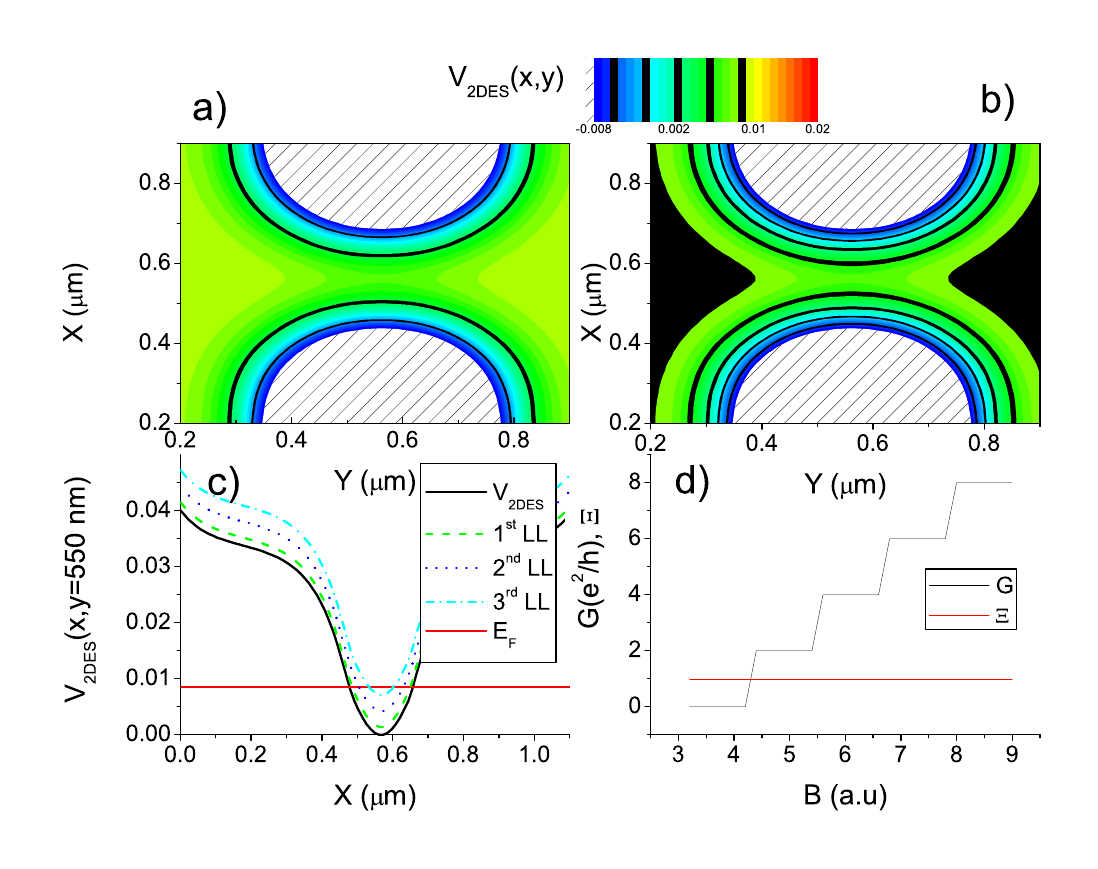}
\caption{ \label{fig:7} The distribution of the spin degenerate
LB-ES at (a) $\nu\approx4$ and (b) $\nu\approx8$. The potential
cross-section at $\nu=8$ plateau (c). Sketch of the conductance
(G) and expected coherence $\Xi$ (d) for C1 considering $W=150$ nm
with an applied gate potential $V_g=2.0$ V, so that the 2DES is
depleted beneath the gates. Since spin degeneracy is assumed each
edge state carries two units of quantized current.}}
\end{figure}
\subsection{Results}
The transport through the QPCs in the absence of an external
magnetic field is well described by the Landauer formalism,
summarized in Eqn.~(\ref{eq:landauer}). The main idea is that the
transport is ballistic and due to the cancellation of the velocity
and a 1D DOS~\cite{Davies}, the conductance is integer multiples
of the conductance unit, $e^2/h$. These integers are just the
number of channels $N_c$, i.e. the number of eigen energies below
the Fermi energy. A similar path is followed when considering an
external $B$ field, which assumes that the conductance is
ballistic within 1D channels and neglects the electron
interactions. These 1D channels are formed whenever the Fermi
energy coincides with a Landau level (Landauer-B\"uttiker
picture). Before proceeding with the full self-consistent solution
of the density and current distribution problem, we first
investigate the positions of the Landauer-B\"uttiker edge states.
The procedure is simple, to obtain the energy dispersion we use
the relation \be E_{n,k_y}(x,y)=E_n+V(x,y)\ee and follow the
equipotential (energy) lines coinciding with the Fermi energy. By
doing so we would be able to discuss qualitatively the phase
coherence $\Xi$, taken to be equal to 1 when there is a full phase
coherence and 0 when there is none. We neglect all the external
sources of decoherence and assume that the LB-ES are coherent at
the length scales we are interested in, i.e. the wave functions of
the associated channels do not overlap.

In Fig.~\ref{fig:7}, we show the spatial distribution of the
expected positions of the LB-ES at two filling factors. The color
scale depicts the self-consistent potential, whereas the black
strips show the LB-ES. The white shaded areas are the electron
depleted regions. The $\nu=2$ and $\nu=4$ edge states nicely show
the expected distribution which are spatially $\sim 40$ nm apart,
Fig.~\ref{fig:7}a. Depending on the steepness of the potential or
the magnetic field value, however, this distance may become less.
For the etched samples (not shown) at the same filling factor the
spatial distance between $\nu=2$ and $\nu=4$ edge states become
almost half the value of the gated samples. For a filling factor
$\nu=$8 plateau the outermost (the ones closest to the gates) two
edge states are less than 15 nm apart from each other, and the
wave functions extend over a larger distance. It is apparent that,
when the two wave functions start to overlap, the coherence $\Xi$
is reduced drastically. However, for the ideal case (no overlap)
$\Xi$ should stay constant for all plateau regions, since by
definition the edge states can not cross each other. The
conductance quantization is, of course, independent of the
structure of the ES itself and according to
Eqn.~(\ref{eq:landauer}) one should simply count the number of ES
which cross the constriction. The conductance is shown by the
sketch in Fig.~\ref{fig:7}d, of course the sharp transition
between the plateaus is changed when one considers level
broadening or in general scattering. One should note that,
although the LBES picture is useful in making qualitative
arguments, one needs to grasp the actual distribution of the edge
states to understand the physics observed at the
experiments~\cite{Litvin:2008arXiv}.

Next we investigate the distribution of the incompressible strips
calculated self-consistently described by
Eqns.~(\ref{eq:tfadensity}-\ref{eq:tfapotential}). The conductance
through the QPC can be rewritten \be G=\frac{e^2}{h}\nu_{\rm
center} \ee as conjectured in
Ref.~[\onlinecite{Chklovskii93:12605}], where $\nu_{\rm center}$
is the filling factor at the very center of the QPC. It is
apparent that, if this value is an integer, i.e. incompressible,
the conductance is quantized. Therefore, it is important to study
this condition for a realistic QPC geometry. In the following we
first calculate the filling factor distribution in the absence of
an external current and then obtain the current distribution in
the next section.

\begin{figure}
{\centering
\includegraphics[width=1.\linewidth]{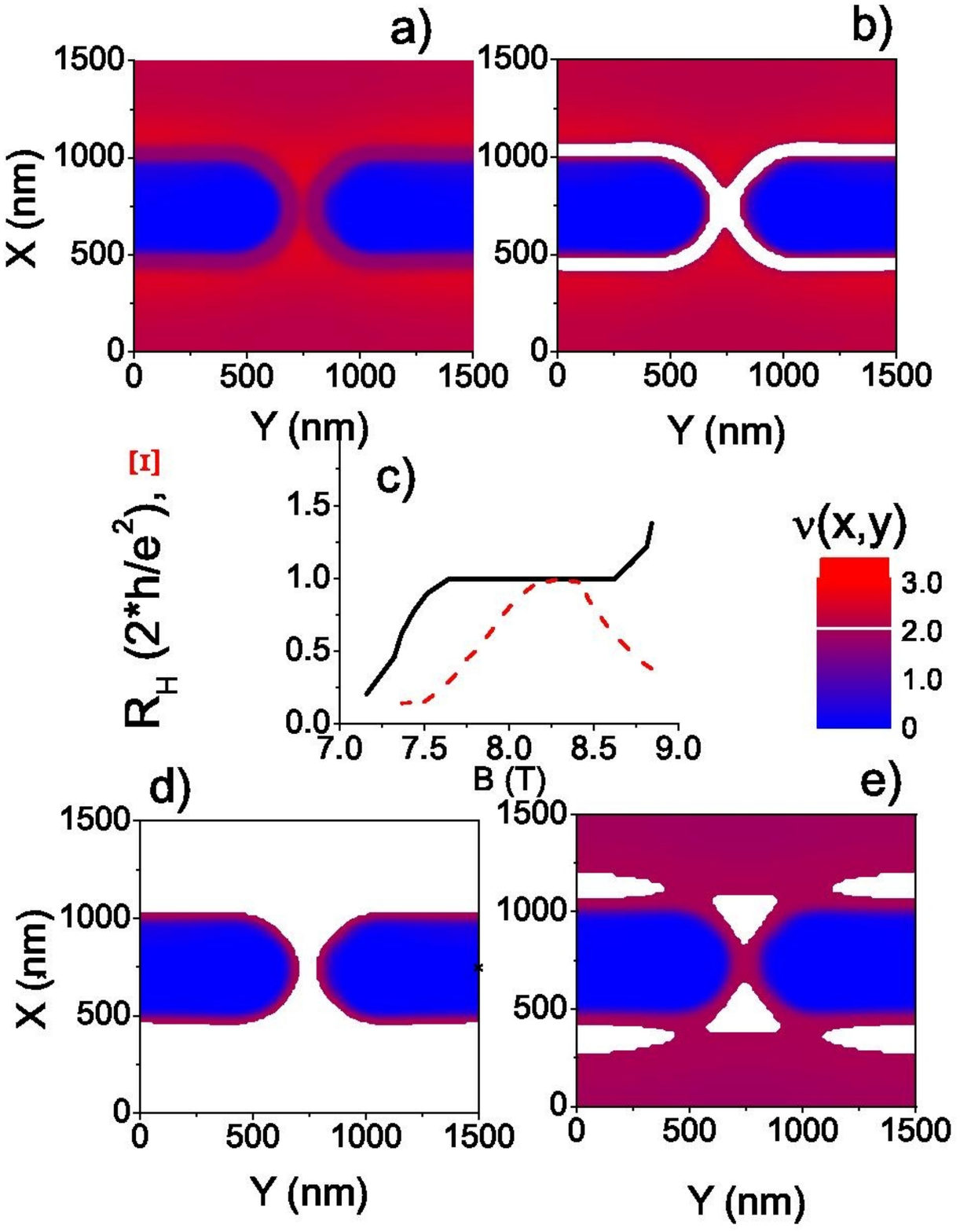}
\caption{ \label{fig:8} Spatial distribution if the incompressible
strips (white areas) for characteristic $B$ values (a) 6.8 T, (b)
7.3 T, (d) 8.3 T (e) 8.8 T calculated at temperatures
$\hbar\omega_c/k_BT\ll 1$, together with the sketch of Hall
resistance and the coherence $\Xi$ (c). The QPC configuration
considered here is C1 with $W=150$ nm, the gate voltage is chosen
such that all the electrons beneath the gates are depleted.}}
\end{figure}

In order to cure the artifacts arising from TFA i) we consider a
DOS broadened by a Gaussian~\cite{Guven03:115327}given by \be
D(E)=\frac{1}{2\pi
l^2}\sum_{n=0}^{\infty}\frac{\exp(-[E_n-E]^2/\Gamma_{\rm
imp}^2)}{\sqrt{\pi}\,\Gamma_{\rm imp}} \ee with the impurity
parameter $\Gamma_{\rm imp}$, which is chosen large enough
$\Gamma_{\rm imp}/\hbar \omega_c=0.3$ to cover the level
broadening generated by the local electric
field~\cite{Kramer03:172} and self-consistent broadening
effects~\cite{Cai86:3967} and ii) a spatial averaging is carried
out over the Fermi wavelength ($\sim30$ nm). Fig.~\ref{fig:8}
summarizes our results showing the spatial distribution of the
incompressible ES, considering the quantum Hall plateau $\nu=2$.
Pedagogically, starting our investigation from large magnetic
fields is preferable; at large magnetic fields
(Fig.~\ref{fig:8}e), the system is mostly compressible (colored
area) and the two incompressible (white) regions do not merge at
the opening of the QPC. Therefore, both the Hall resistance and
the conductance through the QPC is not quantized. As soon as one
enters to the QH plateau almost all of the sample becomes
incompressible shown in Fig.~\ref{fig:8}d (in the absence of short
range impurities) and both $R_H$ and $G$ becomes quantized.
Decreasing the magnetic field creates two incompressible ES which
are spatially separated seen in Fig.~\ref{fig:8}b, however, the
quantization is not affected. At a lower magnetic field value
these IS-ES disappear (Fig.~\ref{fig:8}a) as an end result of
level broadening and (simulation) of the finite extend of the wave
functions, now we are out of the QH plateau and G is no longer
quantized. This picture and the LB-ES picture yield same behavior
for the $R_H$ and $G$, however, in the later one current is
carried by the IS-ES, which we will discuss in the next section.
The qualitative difference between the two pictures is the
coherence as shown by the (red) dashed line in Fig.~\ref{fig:8}c.
First $\Xi$ presents minima in between two plateau regimes, since
the IS die out, second at the higher edge of the QH plateau the
system becomes completely incompressible therefore, it is not
possible to define separate ESs hence coherence is lost
(averaged). We believe that, this nonuniform behavior of the
coherence within the QH plateau coincides with the experimental
findings of Roche et.al.~\cite{Roche07:QHP}, however, we admit
that other explanations are also possible. Another interesting
experimental work is carried by the Regensburg group, where they
have investigated the amplitude of the visibility oscillations as
a function of $B$ field~\cite{Litvin:2008arXiv}. They have
reported a maximum visibility at $\nu\thickapprox 1.5$, which
seems quite opposite to, what has been reported by other
groups~\cite{Heiblum03:415,Roche07:QHP}. However, it is easy to
see that their sample has a homogeneous width all over, which is
not the case for other groups. From self-consistent
calculations~\cite{Siddiki:ijmp} it known that, if the sample
width is narrower than 5-6 $\mu$m the center electron density (or
filling factor) can differ strongly from that of the average
one(s). An indication of such a case is also shown by numerical
simulations~\cite{ahmet}.
In interconnecting the Hall plateaus and the spatial distribution
of the incompressible strips, we have used the findings of
Ref.~[\onlinecite{siddiki2004}] where the current is shown to be
flowing through the incompressible strips. This is in contrast to
some of the models~\cite{Chklovskii92:4026,Beenakker89:2020} in
the literature where the opposite is proposed. In the next
section, we will present the general concepts of the local Ohm's
law and based on the absence of back scattering in the
incompressible strips we will show that the local resistivity
vanishes and the external current should be confined to these
regions.
\section{Current distribution within local Ohm's law\label{sec:4}}
The local (potential) probe
experiments~\cite{Ahlswede01:562,Ahlswede02:165,Yacoby00:3133},
brought novel information concerning the Hall potential
distribution over the sample. The first set of experiments show
clearly that the potential, therefore the current, distribution is
strongly magnetic field dependent. It was shown that, out of the
QH plateau regime the Hall potential varies linearly (Type I) all
over the sample, a similar behavior to classical (Drude) result.
Whereas, at the lower edge of the QH plateau the current is
confined to the edges of the sample (Type II), which was shown to
be coinciding with the positions of the incompressible strips. The
most interesting case is observed when an exact (even) integer
filling is approached. In these magnetic field values, the
potential exhibits a strong nonlinear variation all over the
sample, which was attributed to the existence of a large (bulk)
incompressible region. The explanation of these measurements
acquired a local transport theory, where the conductivities and
therefore current distribution can be provided also taking into
account interactions. In the subsequent theoretical
works~\cite{Guven03:115327,siddiki2004,Siddiki04:condmat} the
required conditions were satisfied and an excellent agreement with
the experiments were obtained~\cite{Ahlswede01:562}. In the second
set of experiments~\cite{Yacoby04:328} a single electron
transistor has been placed on top of the 2DES and the local
transparency, i.e. whether the system is compressible or
incompressible, and the local resistivity have been measured.
Comparing the transparency and the longitudinal resistivity, it
has been concluded that the resistivity vanishes when the system
is incompressible.

Theoretically, if the local electrostatic potential and the
resistivity tensor $\hat{\rho}(x,y)$ are known the current
distribution $\vec{j}(x,y)$ can be obtained from the local version
of the Ohm's law \be \vec{E}(x,y)= \hat{\rho}(x,y).\vec{j}(x,y)
\label{eq:ohm}\ee provided that \be \vec{E}(x,y)=\nabla
\mu^*(x,y)/e \label{eq:chem}\ee where the electrochemical
potential is position dependent when an fixed external current
$I=\int_A{\vec{j}(x,y)dxdy}$ is imposed. In our calculations we
assume a local equilibrium in order to describe the stationary
non-equilibrium state generated by the imposed current, starting
from a thermal equilibrium state obtained from the modified TFA.
At this point if the local resistivity tensor is known, Eqn.s
(\ref{eq:ohm})-(\ref{eq:chem}) should be solved once again
iteratively for a given electron density and potential profile,
where the equation of continuity \be
\bf{\nabla}\cdot\textbf{j}(\textbf{r})=0 \label{eq:eqofcont}\ee
also holds. We assume that the local resistivity is related to the
local electron density via the conductivity tensor, i.e.
$\hat{\rho}(x,y)=\hat{\sigma}^{-1}(n_{\rm el} (x,y))$. For a
Gaussian broadened DOS the longitudinal component of the
conductivity tensor is obtained from \be
\sigma_l=\frac{2e^2}{h}\int_{-\infty}^{\infty}dE\large[-
\frac{\partial f}{\partial E}\large]\sum_{n=0}^{
\infty}\large(n+\frac{1}{2}\large) [e^{(-[\frac{E_n-E}{\Gamma_{\rm
imp}}]^2)}] \label{eq:sigmaL}\ee whereas Hall component is
simply\be \sigma_H=\frac{2e^2}{h}\nu \label{eq:sigmaH},\ee where
we ignored the self-energy corrections depending on the type of
the impurity scatterers. We should emphasize that, the above
conductivities are used for consistency reasons, in principle, any
other reasonable impurity model like the commonly used
SCBA~\cite{Ando82:437,siddiki2004} can be considered. Assuming
that TFA is valid, we can replace the local conductivities with
the above defined global ones.

\begin{figure}
{\centering
\includegraphics[width=1.\linewidth]{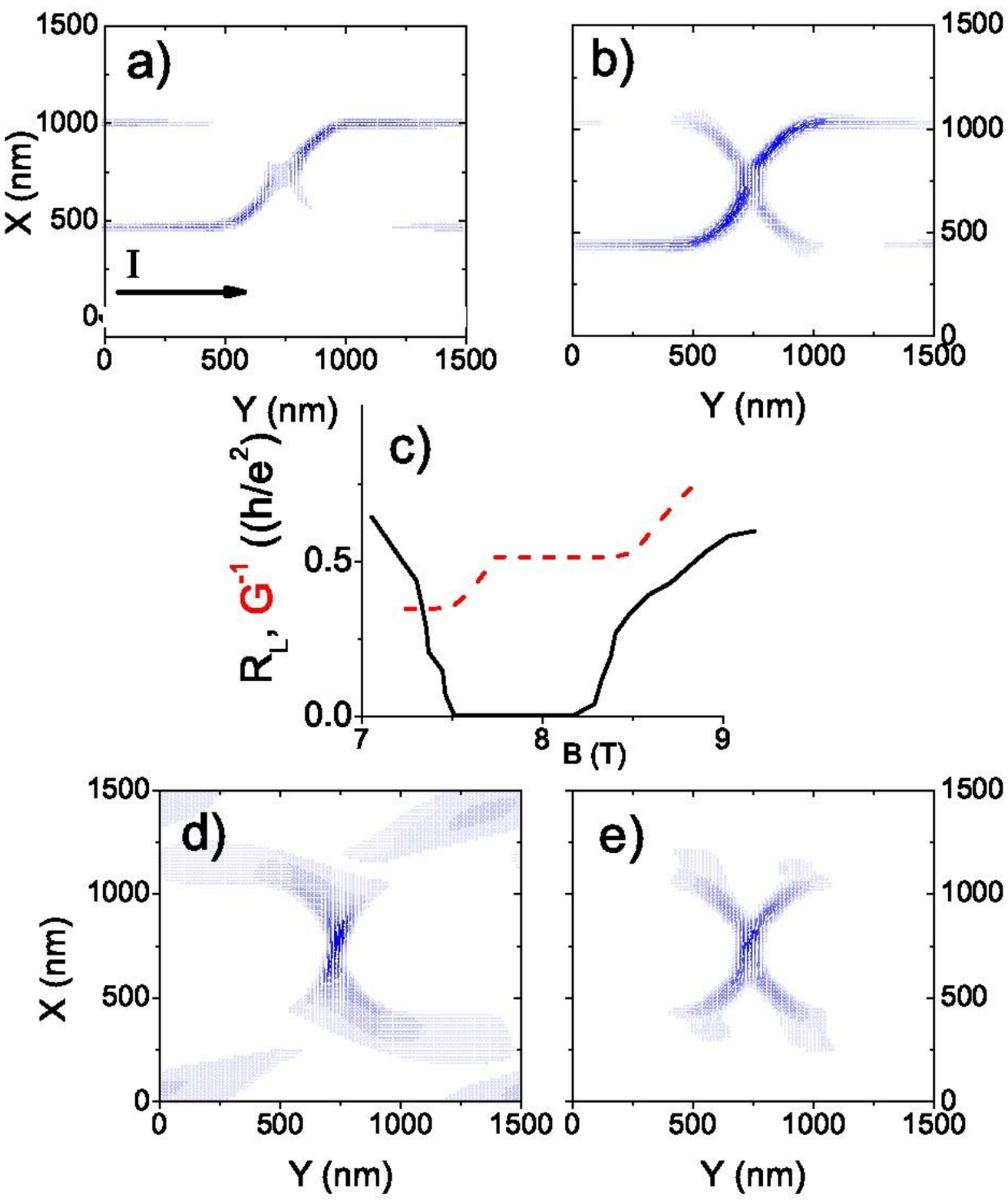}
\caption{ \label{fig:9} The local current density calculated at
different field strengths, same as in Fig.\ref{fig:8}. The
intensity of the current density is chosen such that, the applied
current does not effect the density distribution, i.e.
$|j(x,y)|\sim 0.4\times 10^{-3}$ A/m.}}
\end{figure}
In the absence of an external current our calculation scheme is as
follows, we initialize Eqn.~(\ref{eq:tfadensity}) using the total
potential obtained from the 3D calculations and obtain the
electron density at relatively high temperatures ($kT/\hbar
\omega_c \sim 0.5$) and use this density distribution to obtain
resulting potential from Eqn.~(\ref{eq:tfapotential}). Next we
keep on iterating until a numerical convergence is reached, where
the electron density is kept constant. This is followed by the
step where the temperature is lowered by a small amount and
iteration process is repeated till the target temperature is
reached. After the thermal equilibrium is obtained, we impose a
small amount of external current and solve
Eqns.~(\ref{eq:ohm}-\ref{eq:chem}) self-consistently. While doing
this second iteration, we fix the constant arising from the
integral equation to a value such that the total number of
electrons is kept constant with and without current. As a
numerical remark, if the current loop does not converge we
increase the temperature by a relevant amount and start the
iteration procedure. The whole calculation scheme is composed of
three different codes, which are written in C++, Fortran and
Matlab respectively. In order to obtain reasonable grid resolution
parallel computation techniques were used~\cite{Sefa:Diss}.

Since we are interested in current distribution and also its
effect on the density distribution we find appropriate to present
our results in two separate sections i) where the applied current
is weak enough that the electron and potential distribution is
unaffected, linear response. ii) the applied current is
sufficiently large so that the imposed current induces a
considerable change on the position dependent electrochemical
potential, out of linear response.
\subsection{Linear response regime}
The crucial part of the local approach is that for a given (large)
magnetic field we can calculate the local potentials and electron
distributions self-consistently. The result of such a calculation
is that the 2DES is essentially separated in two regions, i.e.
compressible and incompressible, therefore for a given (obtained)
density we can calculate the local conductivities via
Eqns.~(\ref{eq:sigmaL}-\ref{eq:sigmaH}). Let us now discuss the
distinguishing conductance properties of these two regions
starting from a compressible region. At a compressible region the
Fermi energy is pinned to one of the Landau levels, screening is
nearly perfect, self-consistent potential is flat and filling
factor is locally a non-integer. According to
Eqn.(\ref{eq:sigmaH}), the Hall conductivity is a non-integer and
the longitudinal conductivity is non-zero meaning finite back
scattering. Now the classically defined drift velocity, and also
its quantum mechanical counter part, is proportional to the
electric field perpendicular to the current direction. We have
seen that at the compressible region the potential perpendicular
to the current direction is flat, therefore the $x$ component of
the electric field is zero, hence the drift velocity. Meanwhile,
at an incompressible region the Fermi energy is in between two
Landau levels, the filling factor is fixed to an integer value and
potential presents a variation perpendicular to the current
direction. Due to the Landau gap the longitudinal conductivity
vanishes, whereas the Hall conductivity assumes its (quantized)
integer value. If one calculates the inverse of the conductivity
tensor for the longitudinal component one obtains, \be
\rho_l(x,y)=\frac{\sigma_l(x,y)}{\sigma^2_l(x,y)+\sigma^2_H(x,y)}
\ee thus the longitudinal resistivity vanishes within the
incompressible region pointing the absence of back scattering. Of
course, the simultaneous vanishing of both the longitudinal
resistivity and the conductivity is a result of applied external
(and perpendicular) $B$ field and is obtained only in two
dimensions. Moreover, since the $x$ component of the electric
field is now non-zero, the drift velocity is finite and the
current is confined to this region. Combining these two one
concludes that, if there exists an incompressible region somewhere
in the sample all the external current is confined to this region
otherwise (if there are no incompressible regions and all the
system is compressible) the current is distributed according to
Drude formalism, i.e. the current density is proportional to the
electron density.

We start our discussion of the current distribution when a small
current is imposed for which the electrostatic and electrochemical
potential satisfies the linear response relation \be
V((x,y);I)-V((x,y);0)\approx\mu^*((x,y);I)-\mu^*_{\rm eq}
\label{eq:linearrespone}.\ee This condition essentially states
that, the imposed current does not modify the electrochemical
potential therefore the electron density remains unchanged.
Fig.~\ref{fig:9} presents the current distribution which is
calculated for the density distribution shown in Fig.~\ref{fig:8}.
The correspondence between the positions of the incompressible
strips and the current density is one to one. In the out of
plateau regime the current is essentially distributed all over the
sample, where no incompressible regions exist, Fig.\ref{fig:9}a.
As soon as one enters the QHP, i.e. when a large \emph{bulk}
incompressible strip (region) is formed, the essential future of
current distribution is not effected strongly, however, in this
situation current is flowing in the incompressible region.
Tracking the positions of the ISs in Fig.~\ref{fig:8}b, we can
readily guess the distribution of the current density in
Fig.~\ref{fig:9}b. Following our arguments about the smearing out
of the narrow incompressible strips, we have a situation in which,
again, the current is spread over the sample shown in
Fig.~\ref{fig:9}a. Although, the IS vanishes the reminiscence of
it still provides a narrow strip of small longitudinal resistivity
and, therefore a higher amount of current is kept confined to
these regions. Fig.~\ref{fig:9}c, presents the corresponding
longitudinal resistance, when measured as a function of $B$
together with the conductance across the QPC. The relation between
$G$ and $R_L$ is interesting, the conductance is quantized, as
soon as $R_L$ vanishes at large fields, however, becomes
non-quantized even though the $R_L=0$ at the lower edge of the
zero resistance state. Let us first discuss the $B$ dependence of
the $R_L$, it is finite if the system is compressible and is zero
if an incompressible strip percolates from one edge of the sample
to the other edge in the current direction, i.e. from source to
drain. Therefore, existence of an IS percolating is sufficient
enough to measure zero longitudinal resistance. However, to have a
conductance quantization the center of the QPC should be
incompressible, which is a stronger
restriction~\cite{Chklovskii93:12605}. Hence, in the lower edge of
the QHP, the IS percolates but the center of the QPC is
compressible. The implication of this fact to the coherence is a
bit more complicated, we have seen that as soon as one enters to
the QHP a large bulk IS is formed therefore the phase of the
electrons is highly averaged. This implies that the coherence is
relatively less than that of the two well separated ISs. On the
other hand, at the lower edge of the conductance plateau, the ISs
become narrower and are less immune to decoherence effects arising
from the environment, hence, the coherence is reduced. Our above
discussion coincides with the recent experiments performed in
small Mach-Zehnder interference devices (MZI), where the
visibility is measured as a function of the external magnetic
field~\cite{Roche07:QHP}. It is fair to note that, some other
mechanisms providing $B$ dependence of decoherence can also
account for such a behavior.

In the mentioned MZI
experiments~\cite{Heiblum03:415,Neder06:016804} and also the
measurements performed at the group of S.
Roddaro~\cite{Roddaro05:156804} a finite (and large) source drain
voltage $V_{\rm SD}$ is applied either to measure the $V_{\rm SD}$
dependency of the visibility or the transmission. The intensity of
the applied current, in these experiments, can not be treated
within the linear response regime, where the electrochemical
potential remains constant, i.e. position independent. In the next
section, we present the current and the density distribution
calculated where Eqn.~(\ref{eq:linearrespone}) does not hold any
more.
\subsection{Beyond linear response}
In the absence of an external current an equilibrium state is
obtained by solving the Eqns.~(\ref{eq:tfadensity}) and
(\ref{eq:tfapotential}) self-consistently. Even in the presence of
a small current, a Hall potential is generated which, in
principle, modifies the electrochemical potential, i.e. tilts the
Landau levels. This modification can be compensated by the
redistribution of the electrons, which certainly modifies the
total electrostatic potential. If the applied current is
sufficiently small, the modification is negligible, i.e. linear
response. However, if the current is large, the resulting Hall
potential is also large and one should re-calculate the electron
density, and therefore the potential distribution till a steady
state is reached. In this section we present the current and
density distribution in the presence of a large external current,
where a local equilibrium is assumed implicitly.
\begin{figure}
{\centering
\includegraphics[width=1.\linewidth]{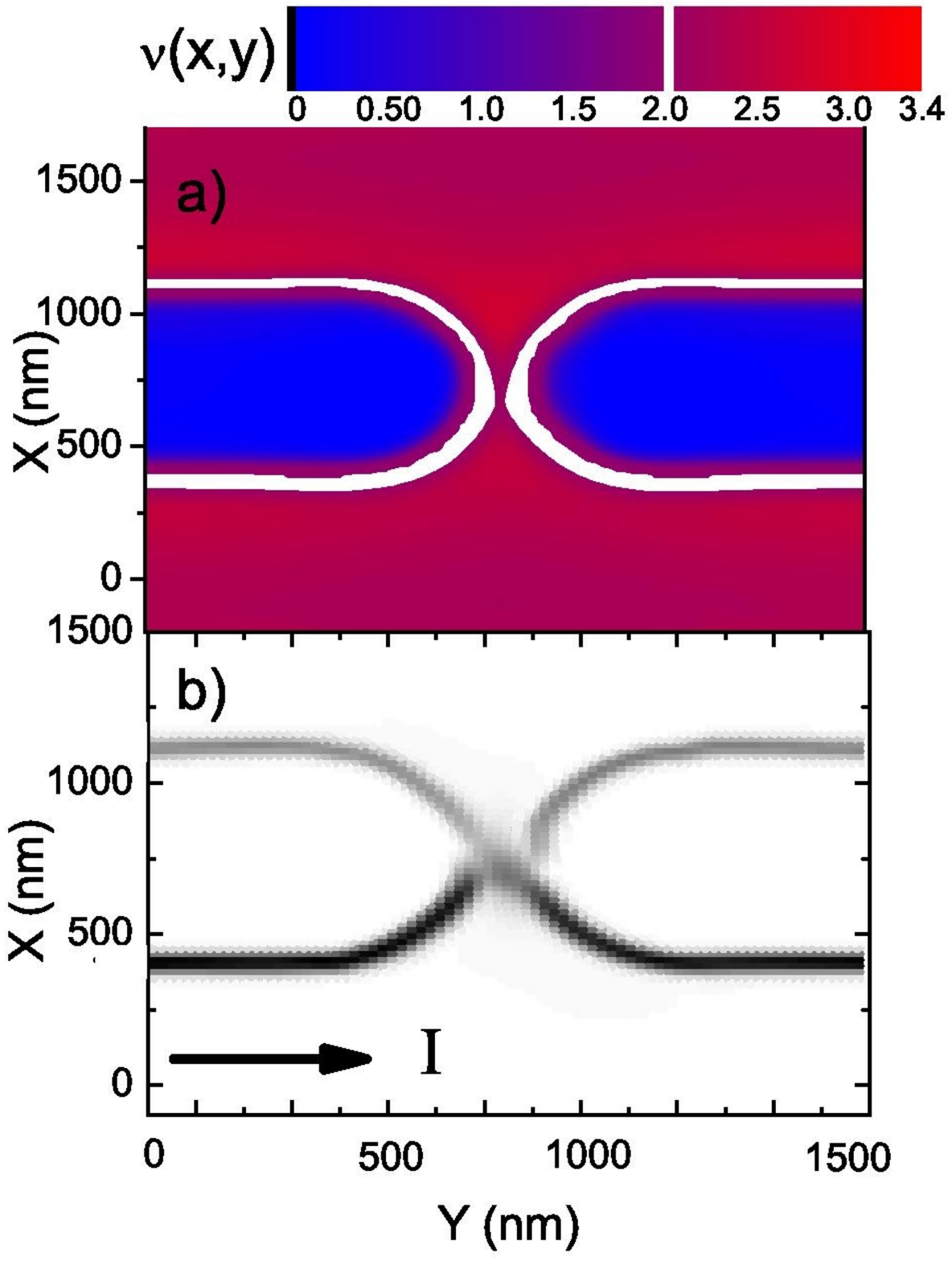}
\caption{ \label{fig:10} The local filling factor distribution and
the corresponding local current density calculated at the default
temperature. The current intensity is sufficiently high
($|j(x,y)|=2.0\times 10^{-2}$ A/m) to induce an asymmetry on the
density distribution via position dependent electrochemical
potential. }}
\end{figure}
Fig.~\ref{fig:10}a shows the electron density distribution in
color scale for $B=7.1$ T, where the intensity of the applied
current is in the out of linear response regime. Note that the $B$
value is chosen such that the Hall resistance is quantized,
however, the conductance is not. The general behavior is similar
to that of linear response, however, it is clearly seen that the
widths of the ISs are asymmetric with respect to $y=750$ nm line,
where current is driven in $y$ direction. The asymmetry is induced
by the large current, since the electrons are redistributed
according to the new (self-consistent) potential distribution. The
corresponding current density distribution is plotted in
Fig.~\ref{fig:10}b, once more the one to one correspondence
between the positions of the ISs and the local current maxima is
apparent. The consequence of the asymmetry and thereby the
widening of the ISs can be observed in the conductance and the
$R_L$, such that the narrow IS at the upper side is smeared out
much earlier than the one on the lower side. We should note that,
such large currents heat the sample therefore the local
temperature within the ISs is larger compared to the lattice
temperature due to Joule heating~\cite{Akera01:1468}. Such a
(local) temperature dependence is treated explicitly by H.
Akera~\cite{Akera06:} and his co-workers and a strong evidence is
provided towards explaining the breakdown of the IQHE within this
approach. Our present approach lacks such a treatment, therefore
the competition between the widening of the ISs and heating due to
large currents is thought to be more complicated than presented
here. As a consequence, the discussion of the coherence is far
beyond our model, however, we think that the amplitude of the
current when measuring visibility~\cite{Roche07:QHP}] is assumably
still in the linear response regime.

In conclusion, by exploiting the local equilibrium and the
properties of a steady state we have calculated the current
distribution near a QPC in the out of linear response regime. We
have shown that a asymmetry is induced on the density profile due
to the bending of the Landau levels generated by the large Hall
potential. We estimate that, the system can still be considered in
the linear response regime if the 1D current density is smaller
than $0.042\times10^{-2}$ A/m, which certainly depends on the
details of the sample geometry.

\section{Summary}
In this paper, we provided a self-consistent scheme to obtain the
electron density, potential profile and current distribution in
the close vicinity of a QPC, within the Thomas-Fermi
approximation. Starting from a lithographically defined 3D sample,
we calculated the charge distribution at the surface gates, at the
plane of 2DES and for etched samples at the side surfaces. The 3D
self-consistent solution of the Poisson equation enabled us to
present the similarities and differences between an etched and
gate defined QPC. We found that, the relatively deep etched
samples present a sharp potential profile near the edges of the
sample. If the depth of the etching exceeds the depth of the 2DES,
surface charges are calculated explicitly. In the presence of a
quantizing perpendicular magnetic field, we have calculated the
distribution of the incompressible strips as a function of the
field strength. We have argued that, if an incompressible strip
becomes narrower than the magnetic length and/or if the transverse
electric field is sufficiently large, due to Level broadening, the
narrow incompressible strip is smeared. In the next step, the
current distribution is obtained both in the linear response and
out of linear response regimes using a local version of the Ohm's
law assuming a steady state at local equilibrium. It is shown that
the current is confined to the incompressible strips, due to the
absence of back scattering, otherwise is distributed all over the
sample. We have commented on the relation between the existence
and percolation properties of the incompressible strips and the
measured quantities such as the longitudinal resistance and
conductance across the QPC. For the ideal clean sample, i.e. in
the absence of long range fluctuations, it is shown that the QH
plateau extends wider than that of the conductance plateau. In the
out of linear response regime, a current induced density asymmetry
is presented for the first time in such geometries under quantized
Hall conditions. The observable effects of such an asymmetry are
not clarified, since it is also know that large currents increase
the temperature locally due to Joule heating.

A natural extension of the existing model is to include the spin
degree of freedom and thereby exchange and correlation
effects~\cite{Esa07:exchange}. A local spin density functional
theory approach~\cite{Igor07:qpc1,Igor07:qpc2} is the much
promising one among others such as Monte Carlo~\cite{MC} and exact
diagonalization, from computational and application point of view.
In fact, such an approach already exists, however, the current is
handled within the Landauer-B\"uttiker formalism, which we think
is not reasonable in the presence of large incompressible strips.
On the other hand, a time dependent spin density functional model
would be a good candidate to describe current for the geometries
under investigation. The implementation of the Akera's theory,
i.e. Joule heating, to our model is of course desirable which is
been worked presently.

Finally, in order to have a predictive power on the interference
experiments we would like to utilize the existing coherent
transport models in describing the current together with our
electrostatic model, which we are not able at the present. Another
challenge is to simulate the real experimental geometries which
already includes more then a single QPC and contacts etc. The
numerical routine we developed is now able to do such large scale
calculations within the linear response regime, however, still
lacks describing the exchange and correlation effects.

\textbf{Acknowledgement}

The authors would like to thank R.R. Gerhardts for his initiation,
supervision and contribution in developing the model. The
enlightening discussions with V. Golovach and L. Litvin are also
highly appreciated. The authors appreciate financial support from
SFB 631, NIM Area A, T\"UBITAK Grant No. 105T110, and Trakya
University research fund under project No. T\"UBAP- 739-754-759.


\end{document}